\documentclass[runningheads]{llncs}
\usepackage{graphicx}
\usepackage{listings, silver}
\usepackage{color, colortbl}
\usepackage{xspace}
\usepackage{cite}
\usepackage{tikz}
\usepackage[normalem]{ulem} %% Crossing out text (defines \sout)

\usepackage[inline]{enumitem} %% Provides inline lists, e.g. environment enumerate*
  \setlist*[enumerate]{label=(\arabic*)}

\usepackage{amsmath}
\usepackage{qsymbols}
\usepackage{stmaryrd} %% E.g. for \Mapsto
\usepackage{multirow}

\usepackage{suffix} %% Be able to declare starred macros, e.g. \foo*
\usepackage{xparse}
\usepackage[colorlinks,linkcolor=black,citecolor=blue]{hyperref}
\usepackage{tabularx}

\newcommand*{\ListingsFontSize}{\fontsize{8.0}{8.3}\selectfont}
\newcommand*{\SilFont}{\ttfamily\ListingsFontSize}
\usepackage{silver}

\definecolor{light-gray}{gray}{0.9}
\definecolor{darkgreen}{rgb}{0.0,0.7,0.0}
\definecolor{darkred}{rgb}{0.55, 0.0, 0.0}
\definecolor{persianblue}{rgb}{0.11, 0.22, 0.73}
\definecolor{navyblue}{rgb}{0.0, 0.0, 0.5}
\definecolor{darkpowderblue}{rgb}{0.0, 0.2, 0.6}
\definecolor{frenchblue}{rgb}{0.0, 0.45, 0.73}
\definecolor{burntorange}{rgb}{0.8, 0.33, 0.0}

\definecolor{rowShade}{gray}{0.85}

\newcommand*{\Figref}[1]{Fig.~\ref{fig:#1}}
\newcommand*{\figref}{\Figref}
\newcommand*{\Secref}[1]{Sec.~\ref{sec:#1}}
\newcommand*{\secref}{\Secref}

\newcommand*{\lineref}[1]{line~\ref{line:#1}}

\newcommand*{\linerange}[2]{lines~\ref{line:#1}-\ref{line:#2}}
\newcommand*{\Sref}[1]{\hyperref[#1]{\S\ref*{#1}}}

\newcommand*{\ie}{i.e.\xspace}

\newcommand*{\iip}{implementation proof\xspace}

\newcommand*{\termacronym}{}%
\NewDocumentCommand{\term}{O{} m}{%
  \textit{#2}%
  \renewcommand*{\termacronym}{}%
  \ifdefmacro{#1}{%
    \ifdefvoid{#1}{}{\renewcommand*{\termacronym}{#1}}%
  }{%
    \ifthenelse{\equal{#1}{}}{}{\renewcommand*{\termacronym}{#1}}%
  }%
  \ifdefvoid{\termacronym}{}{ (\termacronym)}%
  \xspace%
}

\newcommand*{\GobraFont}{\SilFont}
\newcommand*{\GobraFontSize}{\ListingsFontSize}
\newcommand*{\InlineGobraFontSize}{\small}

\lstdefinelanguage{gobra}{
  language=go,
  sensitive=true,
  morecomment=[l]{//},
  morecomment=[s]{/*}{*/},
  morekeywords=[1]{ %% Keywords of the programming language subset
    pred, implements
  },
  morekeywords=[2]{ %% Keywords of the specification language subset
    requires, ensures, invariant, req, ens, pure
  },
  morekeywords=[3]{ %% Keywords of the proof language subset
    fold, unfold, unfolding, in, ghost,
    assume, assert, inhale, exhale
  },
  basicstyle={\GobraFont\GobraFontSize},
  commentstyle={\color[HTML]{747678}\textit},
  keywordstyle={[1]\color[HTML]{0005FF}},%\bfseries
  keywordstyle={[2]\color{burntorange}},%\bfseries
  keywordstyle={[3]\color[HTML]{EC008C}},%\bfseries
  mathescape=true,
  escapechar=§,
  moredelim=**[is][\normalfont\itshape]{'}{'} 
}

\lstdefinelanguage{myViper}{
  language=silver,
  sensitive=true,
  basicstyle={\GobraFont\GobraFontSize},
  mathescape=true,
  escapechar=§,
  moredelim=**[is][\normalfont\itshape]{'}{'} 
}

\lstdefinelanguage{plaintext}{
  language=,
  keywordstyle=\bfseries,
  ndkeywordstyle=\bfseries,
  morecomment=[l]{//},
  mathescape=true
}

\makeatletter
\lst@Key{countblanklines}{true}[t]%
    {\lstKV@SetIf{#1}\lst@ifcountblanklines}

\lst@AddToHook{OnEmptyLine}{%
    \lst@ifnumberblanklines\else%
       \lst@ifcountblanklines\else%
         \advance\c@lstnumber-\@ne\relax%
       \fi%
    \fi}
\makeatother

\lstnewenvironment{gobra}[1][]{
  \lstset{
    language=gobra,
    floatplacement={tbp},
    captionpos=b,
    frame=lines,
    numbers=left,
    numberblanklines=false,
    xleftmargin=8pt,
    xrightmargin=8pt,
    countblanklines=false,
    numberstyle=\scriptsize,
    #1
  }
}{}

\lstnewenvironment{myViper}[1][]{
  \lstset{
    language=myViper,
    floatplacement={tbp},
    captionpos=b,
    frame=lines,
    numbers=left,
    numberblanklines=false,
    xleftmargin=8pt,
    xrightmargin=8pt,
    #1
  }
}{}

\newcommand*{\inlgobra}{%
  \lstinline[%
    language=gobra,%
    columns=fixed,%
    basicstyle={\GobraFont\InlineGobraFontSize},%
    mathescape=true,%
    commentstyle={\color{black}\InlineGobraFontSize},
    keywordstyle={[1]\color{black}\InlineGobraFontSize},
    keywordstyle={[2]\color{black}\InlineGobraFontSize},
    keywordstyle={[3]\color{black}\InlineGobraFontSize},
    keywordstyle={[4]\color{black}\InlineGobraFontSize}
  ]%
}
\newcommand*{\gl}{\inlgobra}

\newcommand*{\inlmyviper}{%
  \lstinline[%
    language=myViper,%
    columns=fixed,%
    basicstyle={\GobraFont\InlineGobraFontSize},%
    commentstyle={\color{black}\InlineGobraFontSize},
    keywordstyle={[1]\color{black}\InlineGobraFontSize},
    keywordstyle={[2]\color{black}\InlineGobraFontSize},
    keywordstyle={[3]\color{black}\InlineGobraFontSize},
    keywordstyle={[4]\color{black}\InlineGobraFontSize}
  ]%
}
\newcommand*{\vl}{\inlmyviper}

\usepackage{letltxmacro} %% Provides \LetLtxMacro
\LetLtxMacro\oldttfamily\ttfamily
\DeclareRobustCommand{\ttfamily}{\oldttfamily\InlineGobraFontSize}

\newcommand{\encodeT}[1]{\llbracket #1 \rrbracket}
\newcommand{\shared}[0]{@}
\newcommand{\exclusive}[0]{\bullet}
\newcommand{\encAs}[0]{\triangleq}

\newcommand{\implementationProof}{_{\textsc{Proof}}}

\newcommand*{\Appref}[1]{App.~\ref{sec:#1}}
\newcommand*{\appref}{\Appref}
\usepackage{orcidlink}
\begin{document}
\title{Gobra: Modular Specification and Verification \\ of Go Programs \\ (extended version)}
\titlerunning{Gobra: Modular Specification and Verification of Go Programs}
% If the paper title is too long for the running head, you can set
% an abbreviated paper title here
%
\author{
\mbox{Felix A. Wolf}\inst{1}\orcidlink{0000-0002-8573-2387} \and
\mbox{Linard Arquint}\inst{1}\orcidlink{0000-0002-6230-8014} \and
\mbox{Martin Clochard}\inst{1} \and 
\mbox{Wytse Oortwijn}\inst{2}\orcidlink{0000-0002-5244-2519} \and 
\mbox{Jo{\~{a}}o C. Pereira}\inst{1}\orcidlink{0000-0003-4671-4132} \and
\mbox{Peter M{\"{u}}ller}\inst{1}\orcidlink{0000-0001-7001-2566}
}
\authorrunning{F. A. Wolf et al.}
% First names are abbreviated in the running head.
% If there are more than two authors, 'et al.' is used.
%
\institute{
  Department of Computer Science, ETH Zurich, Switzerland \and
  ESI (TNO), Eindhoven, The Netherlands
  \\
  \email{
    % \{felix.wolf, linard.arquint, martin.clochard\}@inf.ethz.ch, wytse.oortwijn@tno.nl, \{joaocarlos.mendespereira, peter.mueller\}@inf.ethz.ch
    \{felix.wolf, linard.arquint, martin.clochard, joaocarlos.mendespereira, peter.mueller\}@inf.ethz.ch, wytse.oortwijn@tno.nl
  } 
}
\maketitle              % typeset the header of the contribution
\begin{abstract}
Go is an increasingly-popular systems programming language targeting, especially, concurrent and distributed systems. Go differentiates itself from other imperative languages by offering structural subtyping and lightweight concurrency through goroutines with message-passing communication. This combination of features poses interesting challenges for static verification, most prominently the combination of a mutable heap and advanced concurrency primitives.

We present Gobra, a modular, deductive program verifier for Go that proves memory safety, crash safety, data-race freedom, and user-provided specifications. Gobra is based on separation logic and supports a large subset of Go. Its implementation translates an annotated Go program into the Viper intermediate verification language and uses an existing SMT-based verification backend to compute and discharge proof obligations. 
\end{abstract}

\section{Introduction}\label{sec:intro}
Go is an increasingly popular systems programming language targeting, especially, concurrent and distributed systems such as web applications. It combines standard features of imperative languages, such as mutable heap data structures, with less common concepts, such as structural subtyping and lightweight concurrency through goroutines with message-passing communication.

This combination of features poses interesting challenges for static verification, most prominently the combination of a mutable heap and advanced concurrency primitives. Prior research on Go verification handles some of these features, but not their combination. For instance, Lange et al.~\cite{Lange2017, Lange2018} verify safety and liveness of Go's message-passing, but do not consider functional properties about the heap state, whereas Perennial~\cite{Chajed2019} supports heap data structures, but neither channels nor interfaces. 

We present Gobra, an automated, modular verifier for heap-manipulating, concurrent Go programs. Gobra supports a large subset of Go, including Go's interfaces and primitive data structures, both of which have not been fully supported in previous work.
Gobra verifies memory safety, crash safety, data-race freedom, and user-provided specifications. It takes as input a Go program annotated 
with assertions such as pre and postconditions and loop invariants. Verification proceeds by encoding the annotated programs into the intermediate verification language Viper~\cite{MuellerSS16} and then applying an existing SMT-based verifier. In case verification fails, Gobra reports at the level of the Go program which assertions it could not verify.

Gobra's assertion language builds on established concepts:
Gobra uses separation logic style permissions~\cite{Reynolds02} to reason locally about heap data structures. It supports recursive predicates and specification methods to abstract over (possibly unbounded) data structures and their contents. In particular, Gobra has first-class predicates that enable a natural specification of concurrency primitives, for instance, to parameterize a lock by an invariant. 
% Additional annotations make explicit whether data is allocated on the heap or on the stack (which is otherwise hidden by Go's lifetime detection); this distinction is crucial for verification. 

Gobra is intended for the verification of substantial, real-world code, and is currently used to verify the Go implementation of the SCION internet architecture~\cite{zhang2011scion}. Our tool paper makes the following technical contributions:

\begin{enumerate}
\item  We present the Gobra tool, an automated modular verifier for annotated Go programs. Our evaluation demonstrates that Gobra can verify non-trivial examples with good performance. Our artifact is available online~\cite{Artifact}.

\item We define a specification language for functional properties of Go programs. Our specification language provides a consistent abstraction at the level of Go and does not leak details of the underlying encoding.

\item We present the first specification and verification technique for structural subtyping via Go interfaces.
%% Our solution encompasses:
%% \begin{enumerate*}[label=(\alph*)]
%% \item an encoding of part of Go's type system
%% \item a sound treatment of the polymorphic properties of interfaces %% values
%% \item a new specification language for interfaces.
%% \end{enumerate*}
% Go has structural subtyping, which has not been the target of SMT-based verification approaches in the context of imperative programs.

\item Our Viper encoding supports, among other features, Go's broad range of built-in data types, such as slices and channels. A lightweight annotation allows it to apply separation logic to reason soundly about addressable memory locations, but use a more efficient encoding for others.
\end{enumerate}

\subsubsection{Outline.}
We demonstrate key features of Gobra on examples (\secref{nutshell}),
give an overview of the encoding into Viper (\secref{encoding}), 
and provide an experimental evaluation of Gobra (\secref{evaluation}).
Lastly, \secref{conclusion} discusses related work and concludes.

\section{Gobra in a Nutshell}\label{sec:nutshell}
This section illustrates Gobra's specification language on simple examples and shows how we handle interfaces and concurrency.

%%%%%%%%%%%%%%%%%%%%%%%%%%%%%%%%%%%%%%%%%%%%%%%%%%%%%%%%%%%%
\subsection{Basics}
\label{sec:nutshell:basics}
%%%%%%%%%%%%%%%%%%%%%%%%%%%%%%%%%%%%%%%%%%%%%%%%%%%%%%%%%%%%

Gobra uses a variant of separation logic~\cite{Reynolds02} in order to reason about mutable heap data structures and concurrency. Separation logics associate an access permission with each heap location. Access permissions are held by method executions and transferred between methods upon call and return. A method may access a location only if it holds the associated permission. Permission to a shared location \inlgobra{v} is denoted in Gobra by 
\inlgobra{acc(&v)}, which is analogous to separation logic's \inlgobra{v}$\mapsto\_$. Gobra provides an expressive permission model supporting fractional permissions~\cite{boyland03} to allow concurrent read accesses while still ensuring exclusive writes, (recursive) predicates to denote access to unbounded data structures, and quantified permissions (also called iterated separating conjunction) to express permissions to random-access data structures such as arrays and slices.

\begin{figure}[t]
  \makeatletter
  \lst@AddToHook{OnEmptyLine}{\vspace{-0.6\baselineskip}}
  \makeatother
  \begin{gobra}
requires $\forall$ k int $::$ 0 $\leq$ k $<$ len(s) $\implies$ acc(&s[k])
ensures  $\forall$ k int $::$ 0 $\leq$ k $<$ len(s) $\implies$ acc(&s[k])
ensures  $\forall$ k int $::$ 0 $\leq$ k $<$ len(s) $\implies$ s[k] == old(s[k]) + n
func incr (s []int, n int) {
  
  invariant 0 $\leq$ i $\leq$ len(s)
  invariant $\forall$ k int $::$ 0 $\leq$ k $<$ len(s) $\implies$ acc(&s[k])
  invariant $\forall$ k int $::$ i $\leq$ k $<$ len(s) $\implies$ s[k] == old(s[k])
  invariant $\forall$ k int $::$ 0 $\leq$ k $<$ i $\implies$ s[k] == old(s[k]) + n
  for i := 0; i < len(s); i += 1 {
    s[i] = s[i] + n
  }
}
  \end{gobra}
  \caption{A simple Gobra example showing method and loop contracts.}
  \label{fig:basics}
\end{figure}

The example in \figref{basics} illustrates the use of permissions. Method \inlgobra{incr} increases all elements
of a given slice \inlgobra{s} by an amount \inlgobra{n}. 
(Slices are data types that can intuitively be seen as shared arrays of variable length.) The method requires permission to all slice elements (via its precondition) and returns them to the caller (via its first postcondition). 

Functional properties are expressed via standard assertions, which include side-effect free Go expressions (including calls to pure methods, as we explain below) as well as universal quantification and old-expressions to refer to the value an expression had in the pre-state of a method. In our example, the second postcondition uses these assertions to express the functional behavior of the method. The loop invariants are analogous to the method contracts and are needed for verification.

In Go, any memory location can either be \emph{shared} or \emph{exclusive}. Shared locations reside on the heap and can, thus, be accessed by multiple methods and threads; reasoning about shared locations requires permissions to ensure race freedom and to enable framing, \ie, preserving information across heap changes. On the other hand, exclusive locations are accessed exclusively by one method execution and may be allocated on the stack; they can be reasoned about as local variables. The Go compiler determines automatically whether a location is shared or exclusive, for instance by determining whether its address is taken at some point of the execution. To make verification independent of a particular compiler analysis, Gobra requires shared locations to be decorated with an extra annotation \inlgobra{@} at the declaration point, as illustrated by the following client of \inlgobra{incr}:

\begin{gobra}
a@ := [4]int { 1, 2, 4, 8 }
incr(a[2:], 2) $\label{line:basic_array_slice}$
assert a == [4]int { 1, 2, 6, 10 }
\end{gobra}

The first line declares a Go array \inlgobra{a} of fixed length 4, with values 1, 2, 4, and 8. This array is sliced on \lineref{basic_array_slice} using the syntax \inlgobra{a[2:]}, thereby omitting the first two elements of \inlgobra{a} from the created slice. Since \inlgobra{a} is used in a context in which it is sliced, it is a shared location, which is made explicit via the \inlgobra{@} annotation. Consequently, the array creation will produce permissions to the array elements, which are required by \inlgobra{incr}'s precondition. Omitting the \inlgobra{@} annotation will cause a verification error.

%%%%%%%%%%%%%%%%%%%%%%%%%%%%%%%%%%%%%%%%%%%%%%%%%%%%%%%%%%%%%%%%%%%%%%%%
\subsection{Interfaces}
\label{sec:nutshell:interfaces}
%%%%%%%%%%%%%%%%%%%%%%%%%%%%%%%%%%%%%%%%%%%%%%%%%%%%%%%%%%%%%%%%%%%%%%%%

Go supports polymorphism through \emph{interfaces}, named sets of method signatures. Subtyping for interfaces is structural: a type implements an interface iff every method of the interface is implemented by the type. The subtype relationship is determined by the type checker, without any declarations from the programmer\footnote{For the sake of simplicity, we omit \emph{embeddings}, Go's construct for delegation; an extension is straightforward.}.

Calls on an interface value are dynamically dispatched. In settings with nominal subtyping, dynamic dispatch is handled by proving behavioral subtyping~\cite{liskov1994behavioral}: each subtype declaration requires a proof that the specifications of subtype methods refine the specifications of the corresponding supertype methods. Since structural subtypes are not declared explicitly, we adapt this approach as follows.

Whenever a Go program assigns a value to a variable of an interface type, Gobra requires an \emph{implementation proof}, that is, a proof that each method of the subtype satisfies the specification of the corresponding method in the interface. Implementation proofs are inferred automatically by Gobra in simple cases; user-provided implementation proofs are required especially when they include ghost operations, for instance, to manipulate predicates.

\begin{figure}[t]
  \begin{gobra}
  type stream interface{ $\label{line:itf_start}$
    pred memory()
    
    requires acc(memory(), _) // arbitrary fraction of memory() $\label{line:itf_has_next_preconditon}$
    pure hasNext() bool
  
    requires memory() && hasNext()
    ensures  memory()
    next() interface{}
  } $\label{line:itf_end}$
  $\hspace{-10pt}\rule{0.95\textwidth}{0.5pt}$
  type counter struct{ f int; max int } $\label{line:itf_impl_start}$
  
  requires acc(&x.f, _) && acc(&x.max, _)
  pure func (x *counter) hasNext() bool { return x.f < x.max } $\label{line:itf_hasNext}$
  
  requires acc(&x.f) && acc(&x.max, 1/2) && x.hasNext()
  ensures  acc(&x.f) && acc(&x.max, 1/2) && x.f == old(x.f)+1 $\label{line:itf_impl_next_spec_read}$
  ensures  typeOf(y) == int && y.(int) == old(x.f) $\label{line:itf_impl_next_spec}$
  func (x *counter) next() (y interface{}) { x.f++;return x.f-1 }$\label{line:itf_impl_next_body}\label{line:itf_impl_end}$
  
  $\hspace{-10pt}\rule{0.95\textwidth}{0.5pt}$
  pred (x *counter) memory() { acc(&x.f) && acc(&x.max) } $\label{line:itf_proof_start}$

  (*counter) implements stream { $\label{line:itf_proof_clause_start}$
  
    pure (x *counter) hasNext$\implementationProof$() bool { $\label{line:itf_has_next_proof}$
      return unfolding acc(x.memory(), _) in x.hasNext() $\label{line:itf_has_next_proof_body}$
    }
  
    (x *counter) next$\implementationProof$() (res interface{}) { $\dots$ } $\label{line:itf_next_proof}$
  } $\label{line:itf_proof_end}\label{line:itf_proof_clause_end}$
  \end{gobra}
  \caption{An interface specification for a stream (\linerange{itf_start}{itf_end}) together with an implementation (\linerange{itf_impl_start}{itf_impl_end}) and an \iip (\linerange{itf_proof_start}{itf_proof_end}).
We write \gl{acc(p, _)} to denote an arbitrary, positive amount of predicate \gl{p}, and simply \gl{p} for \gl{acc(p, 1/1)}.
At \lineref{itf_impl_next_spec_read}, the fractional permission to \gl{&x.max} entails that \gl{x.max} is not modified.
}
  \label{fig:interfaces}
  \end{figure}

The example in \Figref{interfaces} illustrates this approach.
Interface \gl{stream} (\linerange{itf_start}{itf_end}) declares an interface with two methods, \gl{hasNext} and \gl{next}. The latter may return values of an arbitrary type, which is denoted by an empty interface. Since interfaces do not contain an implementation, their specification must be fully abstract. To this end, \gl{stream} introduces an abstract predicate \gl{memory}, whose definition is provided by the subtypes of the interface. The functional behavior of interface methods can be expressed in terms of pure (that is, side-effect free) abstract methods, here, \gl{hasNext}, which will also be defined in subtypes.

Next, \linerange{itf_impl_start}{itf_impl_end} show an implementation of the interface in the form of a counter.
The counter has a current \gl{f} and maximum \gl{max} value.
As long as the maximum value is not reached, \gl{next} will increase the current value.
At \lineref{itf_impl_next_body}, an integer can be assigned to the empty interface since behavioral subtyping holds trivially.
The specification at \lineref{itf_impl_next_spec} expresses that the returned interface value contains an integer with the old value of the \gl{f} field. 

The counter implementation is completely independent of the \gl{stream} interface. Their connection is established only in the implementation proof (\linerange{itf_proof_start}{itf_proof_end}). This proof defines the \gl{memory} predicate from the \gl{stream} interface for receivers of type \gl{counter} (\lineref{itf_proof_start}). Moreover, an implementation proof verifies that the specification of each method implementation refines the specification of the corresponding interface method. This proof checks that, assuming the precondition of an interface method, a call to the implementation method with identical arguments establishes the postcondition of the interface method. 
This format is enforced syntactically and permits ghost operations before and after the call to manipulate predicates. 
For instance, the proof on \lineref{itf_has_next_proof_body} for \gl{hasNext} temporarily unfolds the \gl{memory} predicate to obtain permission to \gl{x}, which is required by the implementation method, and conversely after the call.

Implementation proofs can be written explicitly, imported from other packages, and also inferred automatically when no explicit proof exists in the current scope. Currently, Gobra does not infer ghost operations such as the \gl{unfolding} on \lineref{itf_has_next_proof_body}; our experiments suggest that already simple heuristics can deal with many cases occurring in practice.
For instance, many implementation proofs we have encountered follow the same pattern: First, the interface predicate instances of the precondition are unfolded. Second, the implementation method is called. Lastly,  the interface predicate instances of the postcondition are folded. This pattern can be generated automatically to alleviate the annotation burden.

Gobra's implementation proofs enable one to reason about interfaces without enforcing subtype declarations in either the interface or the declaration, which would defeat the purpose of structural subtyping. This solution allows one to reason about dynamically-dispatched calls. For instance, the following code snippet verifies in Gobra:

\begin{gobra}
x := &counter{0, 50}
var y stream = x
fold y.memory()
var z interface{} = y.next()
\end{gobra}

In particular, Gobra is able to determine that \gl{next}'s precondition \gl{hasNext()} holds because \gl{y.hasNext()} is equal to \gl{x.hasNext()}, and the latter follows from the definition of \gl{hasNext} (\lineref{itf_hasNext}) and the initial value of \gl{x.f}. This intuitive reasoning is enabled by an intricate underlying encoding, which is not exposed to users.
Users do not have to know how interface predicates are encoded and can treat interface predicates the same as any other separation-logic predicate.

% 2.1.1.1) JML: model fields: cannot be parameterize, have to be maintained, etc
% 2.1.1.2) Verifast: predicate family to abstract over resources and state. You have to manually exchange abstract predicate with 
% 2.1.1.3) Nagini: has no abstract interface, but similar predicate approach. However, not modular (due to reasons that do not occur in Go)
% 2.1.1.4) Vercors: abstract over the state with a single predicate!

% Encoding:
% 1) challenges
% 1.1) modularity

%%%%%%%%%%%%%%%%%%%%%%%%%%%%%%%%%%%%%%%%%%%%%%%%%%%%%%%%%%%%%%%%%%%%
\subsection{Concurrency}
\label{sec:nutshell:concurrency}
%%%%%%%%%%%%%%%%%%%%%%%%%%%%%%%%%%%%%%%%%%%%%%%%%%%%%%%%%%%%%%%%%%%%

Go supports concurrency through \emph{goroutines}, lightweight threads started by prefixing a method call with the \gl{go} keyword. Go offers the usual synchronization primitives, but goroutines idiomatically synchronize via \emph{channels}. Buffered channels provide asynchronous communication, where sending a message blocks only when the buffer is full. Unbuffered channels offer rendez-vouz communication.

Gobra enables verification of concurrent programs by associating 
Go's synchronization primitives with predicates that do not only express properties of data but also express how permissions to shared memory get transferred between threads. For instance, lock invariants may include properties as well as permissions to the data protected by the lock, and channel invariants include properties and permissions of the data sent over a channel. These invariants are specified via ghost operations when the synchronization primitive is initialized.

\begin{figure}[!t]

\begin{gobra}
pred messagePerm(wg *sync.WaitGroup, chunk []int, x, y int) {
  ( $\forall$ i int $::$ 0 $\leq$ i $<$ len(chunk) $\implies$ acc(&chunk[i]) ) && $\dots$ $\label{line:sar_messagePerm_body}$
}

requires $\forall$ i int $::$ 0 $\leq$ i $<$ len(s) $\implies$ acc(&s[i])
func searchAndReplace(s []int, x, y int) {
  var wg@ sync.WaitGroup
  ghost wg.Init()
  c := make(chan []int,4)
  // predicate-name{$\dots$, _, $\dots$} is syntax for partial application
  ghost c.Init(messagePerm{&wg, _, x, y}) $\label{line:sar_channel_init}$
  
  // Spawn workers
  invariant acc(c.RecvChannel(), _)  $\label{line:sar_spawn_worker_start}$
  invariant c.RecvGotPerm() == messagePerm{&wg, _, x, y}
  for i := 0; i < numOfWorkers; i++ { go worker(c, wg, x, y) } $\label{line:sar_spawn_worker_end}$

  // Split slice into chunks, which are sent to workers
  invariant c.SendChannel()
  invariant c.SendGivenPerm() == messagePerm{&wg, _, x, y}
  invariant $\forall$ i int $::$ offset $\leq$ i $<$ len(s) $\implies$ acc(&s[i])
  invariant $\dots$ // constraints on offset and nextOffset
  for offset := 0; offset != len(s); offset = nextOffset {
    nextOffset = $\dots$
    wg.Add(1) $\label{line:sar_add_call}$
    fold messagePerm{&wg, _, x, y}(s[offset:nextOffset])
    c <- s[offset:nextOffset] $\label{line:sar_msg_send}$
  }
  wg.Wait() $\label{line:sar_wait_call}$
}

requires acc(c.RecvChannel(), _) $\label{line:sar_worker_recv_perm}$
requires c.RecvGotPerm() == messagePerm{wg, _, x, y};
func worker(c <- chan []int, wg *sync.WaitGroup, x, y int) {

  invariant acc(c.RecvChannel(), _)
  invariant c.RecvGotPerm() == messagePerm{wg, _, x, y};
  invariant ok $\implies$ messagePerm{wg, _, x, y}(chunk)
  for chunk, ok := <- c; ok; chunk, ok = <-c { $\label{line:sar_msg_received}$
    unfold messagePerm{wg, _, x, y}(chunk)
    $\dots$ // replace x with y in chunk
    wg.Done() // same as wg.Add(-1) $\label{line:sar_done_call}$
  }
}
\end{gobra}
\caption{
    Excerpt showing goroutines, channels, and wait groups.
    The code spawns workers (\lineref{sar_spawn_worker_end}), sends slice chunks through a channel to the workers (\lineref{sar_msg_send}), and then waits on a wait group (\lineref{sar_wait_call}). 
    A worker receives a chunk (\lineref{sar_msg_received}), processes it, and then notifies the wait group (\lineref{sar_done_call}).
    For the sake of simplicity, some details were omitted.
}
\label{fig:channel-perm-comm}
\end{figure}

\Figref{channel-perm-comm} illustrates Gobra's concurrency support using an excerpt from a parallel search-and-replace algorithm (see \appref{full_examples} for the full example). Method \gl{searchAndReplace} spawns a series of worker threads and then sends each of them a chunk of the input slice to process. The worker threads are joined via a wait group \gl{wg}. Method \gl{worker} implements the worker threads.

Gobra associates channels (like \gl{c} in the example) with a predicate to specify properties and permissions of the sent data. The call \gl{c.Init(...)} on \lineref{sar_channel_init} takes this predicate as an argument. As expressed on \lineref{sar_messagePerm_body}, it includes permissions to the chunk a worker operates on. 
For synchronous channels, an additional predicate can specify permissions transferred in the opposite direction, from the receiver to the sender.
Initializing a channel also creates send and receive permissions for the channel, which are used to control which threads may access it. In our example, we transfer a fraction of the receive permission to each worker (\lineref{sar_worker_recv_perm}).

The workers receive permission to the chunk they operate on via a message sent on \lineref{sar_msg_send} and received on \lineref{sar_msg_received}. The transfer back is orchestrated through a wait group, which implements an abstract shared counter. Wait groups are used as follows: The main thread adds to the counter the number of units of work to be done in spawned goroutines (\lineref{sar_add_call}). Each spawned goroutine decreases the counter each time a unit of work is done (via a call to \gl{Done}, \lineref{sar_done_call}). The master can await the counter to reach 0 via a call to \gl{Wait} (\lineref{sar_wait_call}). 
Gobra uses dedicated permissions to express the obligation of a thread to perform units of work before decreasing the counter; each time this happens, permissions are transferred to the wait group and, eventually to the main thread calling \gl{Wait}. We omit the details here for brevity. 

In our example, this mechanism allows the main thread to recover the permissions to the entire slice once the workers have terminated. The example in \Figref{channel-perm-comm} illustrates only the permission aspect of the verification. Functional correctness can be verified easily based on the explained machinery, by specifying a stronger channel invariant that includes the work obligation for each worker. We omit the details here, but see \appref{full_examples} for the complete example.

\section{Encoding}\label{sec:encoding}

Gobra encodes an annotated Go program into a Viper program 
verifying only if the input program is correct.
Many features of Gobra are also present in Viper,
making parts of the encoding straightforward.
For instance, methods, pure methods, and predicates are encoded to their Viper counterpart.
Viper's permission model (including fractions, wildcards, and quantifiers) is similar to Gobra's, but memory is represented differently;
Viper's heap is object-based, where each object contains all declared fields.
Viper's fields store primitive values (including references). To encode Go's compound values such as structs, arrays, slices, and interface values, we use Viper's mechanism to declare mathematical types (such as tuples) using uninterpreted types, uninterpreted functions, and appropriate axioms.
Exclusive Go values are directly represented using these mathematical types. For shared values, there is an indirection via the Viper heap to permit aliasing and apply permission-based reasoning.

\subsubsection{Interfaces.}

As explained in \secref{nutshell:interfaces}, our treatment of Go interfaces relies on interface predicates, specification methods, and implementation proofs. We explain how we handle the former two here; based on this encoding, the encoding of implementation proofs is analogous to methods.

Intuitively, we encode interface predicates as a case split over all possible implementations. All implementations not present in the current scope are subsumed by an abstract default case. Consequently, adding an implementation does not invalidate existing proofs, which enables modular reasoning.
The predicate for the \gl{stream} example (\figref{interfaces}) is encoded as follows:

\begin{gobra}[numbers=none]
predicate memory(x: $\llbracket$interface{}$\rrbracket$) { 
  $\llbracket$typeOf(x) == *counter$\rrbracket$ ? $\llbracket$acc(x.(*counter))$\rrbracket$ : unknownMemory(x)
}
predicate unknownMemory(x: $\llbracket$interface{}$\rrbracket$)

function hasNext(x: $\llbracket$interface{}$\rrbracket$) returns (y: $\llbracket$bool$\rrbracket$)
  req $\llbracket$acc(x.memory(), _)$\rrbracket$ $\label{line:itf_enc_has_next_preconditon}$
  ens $\llbracket$typeOf(x) == *counter$\rrbracket$ $\implies$ y == hasNext$\implementationProof$($\llbracket$x.(*counter)$\rrbracket$)
\end{gobra}

\noindent
The body of the predicate branches on the dynamic type of \gl{x}, with a single case for the (only) given implementation.
The abstract predicate \gl{unknownMemory} encodes the default case. 
The encoding of pure methods such as \gl{hasNext} uses an analogous case split, but uses \gl{hasNext$\implementationProof$}, 
which is part of the implementation proof (\figref{interfaces} \lineref{itf_has_next_proof}) 
and couples the interface and implementation method.
Our encoding of interface predicates is an instance of an \emph{abstract predicate family}~\cite{parkinson2005separation}.
% Abstract predicate families are a flexible mechanism for reasoning about abstraction.
For Go, we have crafted a variant that is well-suited for implementation proofs, pure interface methods, and structural subtyping.

\subsubsection{First-class predicates.}

Our support for concurrency uses first-class predicates, for instance, to specify channel invariants (see \secref{nutshell:concurrency}). We encode first-class predicate values as mathematical types, using defunctionalization. Predicate instances are represented by abstract predicates that take the predicate value as an argument.
First-class predicates enable us to use library stubs to support concurrency primitives such as mutexes and wait groups. These stubs allow us to encode the use of these concurrency primitives via standard method calls. Go's native channel operations are represented analogously.

%We extracted the library specifications from Go's language specification and memory model.

% \section{Architecture}\label{sec:architecture}
% \input{architecture}
\section{Implementation and Evaluation}\label{sec:evaluation}
The Gobra implementation consists of a parser and type checker for annotated Go programs and a translation of those programs into the Viper intermediate verification language. The resulting Viper program is verified using Viper's symbolic execution backend, which in turn uses the Z3 SMT solver~\cite{de2008z3}. Verification errors are translated back to the Go level, such that users are not exposed to the internal encodings.
Users never have to inspect the encoding. Error messages contain the failing assertion and a reason describing why the assertion failed.
Gobra's test suite contains 407~verification tests (with and without errors) with a total of 10'030~LOCs (Go code and annotations)
that take 14.9~minutes to verify.

We evaluated Gobra on 14~interesting verification problems, which include well-known algorithms and data structures, and cover Go's main features, such as interfaces (Examples~7--9)
and concurrency primitives (Examples~13 and~14),
including goroutines, mutexes, wait groups, and channels.
For each example, Gobra verifies memory safety and functional correctness properties. To assess Gobra's performance on failing verifications, we have additionally constructed two incorrect variations of each example, one with a seeded error in the specification and one in the implementation. 

All experiments were executed on a warmed-up JVM on a MacBook Pro with % MacBook is written MacBook (capital "B")
a 2.3~GHz 8-Core Intel Core i9 CPU and 32~GB of RAM, running macOS~11.1 and OpenJDK~11. For each experiment, we measured its verification time using Viper's symbolic execution backend and averaged the duration of twelve executions, excluding the slowest and fastest outlier.

\begin{figure}[t]
	\centering
	\begin{tabularx}{\textwidth}{c | X | c | c | r | r | r} 
 		\# & {Example} & LOC / Spec. & Viper LOC & T [s] & T\textsubscript{spec error} [s] & T\textsubscript{impl error} [s] \\
		 \hline
 		 1  & binary search tree 					& 125 / 140	& 632	& 10.88	& 10.50	& 11.67	\\
 		\rowcolor{rowShade} 2  & dutchflag 							& 22 / 16	& 142	& 2.02	& 1.78	& 1.88	\\
 		 3  & heapsort 							& 47 / 93	& 271	& 16.72	& 19.30	& 15.23	\\
 		\rowcolor{rowShade} 4  & dense and sparse matrix 			& 69 / 62	& 326	& 10.46	& 10.55	& 10.06	\\
 		 5  & binary tree 						& 59 / 20	& 217	& 2.09	& 2.08	& 2.11	\\
 		\rowcolor{rowShade} 6  & running ex. (\Figref{basics})		& 10 / 11	& 164	& 1.71	& 1.70	& 1.70	\\
 		 7  & running ex. (\Figref{interfaces})	& 24 / 16	& 186	& 1.04	& 0.98	& 1.01	\\
 		\rowcolor{rowShade} 8  & list of interfaces					& 46 / 27	& 219	& 1.45	& 1.41	& 1.54	\\
 		 9  & visitor pattern 					& 76 / 30	& 475	& 4.38	& 4.22	& 5.45	\\
 		\rowcolor{rowShade} 10 & zune 								& 31 / 12	& 141	& 1.08	& 1.07	& 1.06	\\
 		 11 & relaxed prefix 						& 25 / 36	& 158	& 7.08	& 5.36	& 4.19	\\
 		\rowcolor{rowShade} 12 & pair insertion sort 				& 50 / 105	& 353	& 15.55	& 12.64	& 13.96	\\
 		 13 & parallel search replace 			& 35 / 94	& 565	& 53.18	& 51.97	& 61.54	\\
 		\rowcolor{rowShade} 14 & parallel sum 						& 31 / 98	& 527	& 58.39	& 50.25	& 57.69	\\
 	\end{tabularx}
  \vspace{4pt}
 	\caption{Experimental results. For each experiment, we list the number of lines of Go code (LOC), number of lines of specification and proof annotations (Spec), and the average verification time in seconds for correct examples (T), errors in the specification (T\textsubscript{spec error}), and errors in the implementation (T\textsubscript{impl error}).
	A line containing both, code and annotations, is counted as one line of Go code and one line of annotation.
	}
 	\label{fig:experiments}
\end{figure}

\figref{experiments} summarizes the results, including the required annotations and verification times for the three variants of each example. The annotation overhead ranges between 0.3~and 3.1~lines of annotations per line of code, which is typical for SMT-based deductive verifiers. Verification times range between a second and a minute per example. The verification times are significantly higher when the verified code uses concurrency features; these examples require quantitatively more and more-complex specifications, which complicates reasoning. Lastly, there is hardly any difference between successful and failed verification attempts. Consistent performance is crucial when verifiers are used interactively, where users run them frequently, especially on programs that do not yet verify.

% \section{Related Work}\label{sec:related-work}
% \input{related_work}
\section{Related Work and Conclusion}\label{sec:conclusion}
Besides Gobra, we are aware of two other verification approaches for Go.
Perennial~\cite{Chajed2019} reasons about concurrent, crash-safe systems. Their core techniques are an extension to the Iris framework~\cite{JungKJBBD18} and independent of Go. They connect their theory to Go programs with Goose, a shallow embedding of Go into Coq~\cite{CoqHomepage}, which proves that Go code complies with a given transition system. In contrast to Gobra, Perennial does not support core Go features such as channels and interfaces.

Several prior works~\cite{Lange2017, Lange2018, gabet2020static} infer behavioral types~\cite{huttel2016foundations} to reason about Go's channel-based message passing.
After they infer behavioral types for a given program, they check safety and liveness properties on the inferred types, using model checkers such as mCRL2~\cite{cranen2013overview}.
Some works use additional analyses to strengthen the provided guarantees.
Lange et~al.~\cite{Lange2018} add a termination analysis to enable one to verify unbounded properties under certain conditions. 
Gabet and Yoshida~\cite{gabet2020static} extend this work by inferring behavioral types on shared variables and locks to additionally reason about data-race freedom, lock safety, and lock liveness.
The approaches by Lange et~al.~\cite{Lange2018} and Gabet and Yoshida~\cite{gabet2020static} are vastly different from Gobra. 
They do not verify code contracts, but instead verify global properties such as deadlock and data-race freedom. Their automation is high and annotation overhead minimal, but their analyses are not modular and do not verify functional properties of code. Furthermore, they do not verify properties about the state of the heap.

There are some prior works that can handle channel-based concurrency and heap-manipulating programs, but these do not apply directly to Go.
Villard et~al.~\cite{villard2009proving} introduce a powerful contract mechanism to specify protocols that channels must adhere to. Their channel specification language is more expressive than the one presented in this paper. Their contracts are finite state machines and thus can have multiple phases. 
However, their channels are always shared between two peers whereas Go supports more advanced concurrency patterns where both channel endpoints are shared between an unbounded number of peers.
Actris~\cite{hinrichsen2019actris, hinrichsen2020actris} is a concurrent separation logic built on top of the Iris framework to reason about session types in an interactive theorem prover.
% We could write something about Gobra aiming for a higher degree of automation.
Actris can go beyond two peers, but to do so, it requires a memory model that is incompatible with Go’s memory model.
Actris models the sharing of channel endpoints via Iris' ghost locks, which to our knowledge, implies sequentialization of sends, and dually receives, which is not guaranteed by Go’s memory model.

Gobra's verification logic and encoding into Viper have been inspired by several other Viper-based verifiers, such as  Nagini~\cite{EilersM18} for Python, Prusti~\cite{AstrauskasMuellerPoliSummers19b} for Rust, and VerCors~\cite{BlomH14} for Java. None of these verifiers address the Go-specific features that Gobra supports. 

\subsubsection{Conclusion.}

We introduced Gobra, the first modular verifier for Go that supports reasoning about a crucial aspect of the language: the combination of channel-based concurrency and heap-manipulating constructs. Moreover, Gobra is the first verifier to support Go's version of interfaces and structural subtyping. In future work, we will expand the properties that can be verified with Gobra, in particular to liveness and hyper-properties. Furthermore, we are applying Gobra to verify the implementation of a full-fledged network router~\cite{zhang2011scion}.
Gobra is hosted on Github at \url{https://github.com/viperproject/gobra}.

\paragraph*{Acknowledgements.}

This project has received funding from the European Union’s Horizon 2020
research and innovation program within the framework of the NGI-POINTER % \textsc{ngi-pointer}
Project funded under grant agreement No 871528.

\newpage
\bibliographystyle{splncs04}
\bibliography{arxiv}

\begin{thebibliography}{10}
\providecommand{\url}[1]{\texttt{#1}}
\providecommand{\urlprefix}{URL }
\providecommand{\doi}[1]{https://doi.org/#1}

\bibitem{AstrauskasMuellerPoliSummers19b}
Astrauskas, V., M\"uller, P., Poli, F., Summers, A.J.: Leveraging {R}ust types
  for modular specification and verification. In: Object-Oriented Programming
  Systems, Languages, and Applications (OOPSLA). vol.~3, pp. 147:1--147:30. ACM
  (2019)

\bibitem{BlomH14}
Blom, S., Huisman, M.: The vercors tool for verification of concurrent
  programs. In: {FM}. Lecture Notes in Computer Science, vol.~8442, pp.
  127--131. Springer (2014)

\bibitem{boyland03}
Boyland, J.: Checking interference with fractional permissions. In: Cousot, R.
  (ed.) {SAS}. Lecture Notes in Computer Science, vol.~2694, pp. 55--72.
  Springer (2003)

\bibitem{Chajed2019}
Chajed, T., Tassarotti, J., Kaashoek, M.F., Zeldovich, N.: Verifying
  concurrent, crash-safe systems with {Perennial}. In: {SOSP}. pp. 243--258.
  {ACM} (2019)

\bibitem{CoqHomepage}
{C}oq consortium, T.: The {C}oq proof assistant, \url{https://coq.inria.fr/}

\bibitem{cranen2013overview}
Cranen, S., Groote, J.F., Keiren, J.J., Stappers, F.P., De~Vink, E.P.,
  Wesselink, W., Willemse, T.A.: An overview of the mcrl2 toolset and its
  recent advances. In: {TACAS}. Lecture Notes in Computer Science, vol.~7795,
  pp. 199--213. Springer (2013)

\bibitem{de2008z3}
De~Moura, L., Bj{\o}rner, N.: {Z3}: An efficient {SMT} solver. In: {TACAS}. pp.
  337--340. Springer (2008)

\bibitem{EilersM18}
Eilers, M., M{\"{u}}ller, P.: Nagini: {A} static verifier for python. In:
  {CAV}. Lecture Notes in Computer Science, vol. 10981, pp. 596--603. Springer
  (2018)

\bibitem{gabet2020static}
Gabet, J., Yoshida, N.: Static race detection and mutex safety and liveness for
  go programs (extended version) (2020)

\bibitem{hinrichsen2019actris}
Hinrichsen, J.K., Bengtson, J., Krebbers, R.: Actris: Session-type based
  reasoning in separation logic. Proc. {ACM} Program. Lang.
  \textbf{4}({POPL}),  1--30 (2019)

\bibitem{hinrichsen2020actris}
Hinrichsen, J.K., Bengtson, J., Krebbers, R.: Actris 2.0: Asynchronous
  session-type based reasoning in separation logic. arXiv preprint
  arXiv:2010.15030  (2020)

\bibitem{huttel2016foundations}
H{\"u}ttel, H., Lanese, I., Vasconcelos, V.T., Caires, L., Carbone, M.,
  Deni{\'e}lou, P.M., Mostrous, D., Padovani, L., Ravara, A., Tuosto, E.,
  et~al.: Foundations of session types and behavioural contracts. {ACM} Comput.
  Surv.  \textbf{49}(1),  1--36 (2016)

\bibitem{JungKJBBD18}
Jung, R., Krebbers, R., Jourdan, J., Bizjak, A., Birkedal, L., Dreyer, D.: Iris
  from the ground up: {A} modular foundation for higher-order concurrent
  separation logic. J. Funct. Program.  \textbf{28}, ~e20 (2018)

\bibitem{Lange2017}
Lange, J., Ng, N., Toninho, B., Yoshida, N.: Fencing off {Go}: liveness and
  safety for channel-based programming pp. 748--761 (2017)

\bibitem{Lange2018}
Lange, J., Ng, N., Toninho, B., Yoshida, N.: A static verification framework
  for message passing in {Go} using behavioural types. In: {ICSE}. pp.
  1137--1148. {ACM} (2018)

\bibitem{liskov1994behavioral}
Liskov, B., Wing, J.M.: A behavioral notion of subtyping. ACM Trans. Program.
  Lang. Syst.  \textbf{16}(6),  1811--1841 (1994)

\bibitem{MuellerSS16}
M{\"{u}}ller, P., Schwerhoff, M., Summers, A.J.: Viper: {A} verification
  infrastructure for permission-based reasoning. In: {VMCAI}. Lecture Notes in
  Computer Science, vol.~9583, pp. 41--62. Springer (2016)

\bibitem{parkinson2005separation}
Parkinson, M., Bierman, G.: Separation logic and abstraction. ACM SIGPLAN
  Notices  \textbf{40}(1),  247--258 (2005)

\bibitem{Reynolds02}
Reynolds, J.C.: Separation logic: {A} logic for shared mutable data structures.
  In: {LICS}. pp. 55--74. {IEEE} Computer Society (2002)

\bibitem{villard2009proving}
Villard, J., Lozes, {\'E}., Calcagno, C.: Proving copyless message passing. In:
  {ASPLAS}. pp. 194--209. Springer (2009)

\bibitem{Artifact}
Wolf, F.A., Arquint, L., Clochard, M., Oortwijn, W., Pereira, J.C.,
  M{\"{u}}ller, P.: {{G}obra: Modular Specification and Verification of Go
  Programs} (2021). \doi{10.5281/zenodo.4716664}

\bibitem{zhang2011scion}
Zhang, X., Hsiao, H.C., Hasker, G., Chan, H., Perrig, A., Andersen, D.G.:
  Scion: Scalability, control, and isolation on next-generation networks. In:
  IEEE Symposium on Security and Privacy. pp. 212--227. IEEE (2011)

\end{thebibliography}

\newpage
\appendix

\section{Extended Discussion of the Encoding}\label{sec:encoding2}

As discussed in \secref{encoding}, Gobra encodes annotated Go programs into Viper programs 
verifying only if the input program is correct.
For this purpose,
Viper provides a simple imperative language,
where a program, as for Gobra, consists of methods, pure methods, and predicates.
%
% big picture
At a high level, % unsurprisingly, 
Gobra encodes methods, pure methods, and predicates
to their Viper counterpart.
Additional Viper members are generated to encode proof obligations and certain operations. 
At a low level,
for the encoding to produce sound results,
an encoding has two key ingredients:
First, we define a \emph{memory encoding} that encodes Go's program state and Gobra's ghost state to a representation in Viper.
Second, we encode operations on Go's program state and Gobra's ghost state into Viper operations preserving the behavior with respect to the memory encoding.
%
% Overview
In this extended discussion, we present more details of the memory encoding.
% and then present some highlights of the encoding of interfaces, first-class predicates, and concurrency primitives.

\subsubsection{Memory encoding.}

% Viper primer
To encode program state, Viper offers several native types, including booleans, integers, and basic mathematical types, such as sets, maps, and sequences. 
For new types, Viper offers a mechanism to declare custom mathematical types (such as tuples) using uninterpreted types, uninterpreted functions, and appropiate axioms.
Viper's heap is object-based. All objects have the single type \vl{Ref} and have all defined fields available. 
With the exception of how memory is represented, Viper's permission model is similar to Gobra's.
Accessing a field \vl{f} of an object \vl{o} requires a permission \vl{acc(o.f)}. 
Viper's support also includes permission fractions, permission wildcards, and quantifiers.

% memory encoding
\newcommand{\typevar}[0]{t}
\newcommand{\otherwisetypevar}[0]{l}
A crucial part of the memory encoding is how Gobra encodes types into Viper.
As stated in \secref{nutshell:basics}, 
Gobra introduces a distinction between shared values, which can be aliased and thus we apply permission based-reasoning,
and exclusive values, which cannot be aliased and are permissionless.
We have augmented the type system to capture this property: For a type $\typevar$, we write $\typevar\exclusive$ and $\typevar\shared$ for an exclusive and shared type, respectively.
Whether something is shared or not matters for the encoding.
Intuitively, shared values are encoded as the memory underlying a data type, whereas exclusive values are encoded as a mathematical object representing the data type itself.
Gobra encodes the most important types as follows:
\begin{gobra}[numbers=none]
$\llbracket$ int$\exclusive$ $\rrbracket$ $\encAs$ Int, $\llbracket$ bool$\exclusive$ $\rrbracket$ $\encAs$ Bool,  $\llbracket$ *$\typevar$$\exclusive$ $\rrbracket$ $\encAs$ $\llbracket$ $\typevar$$\shared$ $\rrbracket$
$\llbracket$ struct{$\typevar_1, \dots, \typevar_n$}$\exclusive$ $\rrbracket$ $\encAs$ Tuple[$\encodeT{\typevar_1\exclusive}, \dots, \encodeT{\typevar_n\exclusive}$],  $\llbracket$ [n]$\typevar$$\exclusive$ $\rrbracket$ $\encAs$ NArray[$\encodeT{\typevar\exclusive}$]  
$\llbracket$ []$\typevar$$\exclusive$ $\rrbracket$ $\encAs$   Slice[$\encodeT{\typevar\shared}$], $\llbracket$ interface{$\dots$}$\exclusive$ $\rrbracket$ $\encAs$ Tuple[Type, Poly]  
$\llbracket$ struct{$\typevar_1, \dots, \typevar_n$}$\shared$ $\rrbracket$ $\encAs$ MemoryTuple[$\encodeT{\typevar_1\shared}, \dots, \encodeT{\typevar_n\shared}$]  
$\llbracket$ [n]$\typevar$$\shared$ $\rrbracket$ $\encAs$  NMemoryArray[$\encodeT{\typevar\shared}$], $\llbracket$ $\otherwisetypevar$$\shared$ $\rrbracket$ $\encAs$  Ref $(\text{with a field of type } \encodeT{\otherwisetypevar\exclusive})$
\end{gobra}
Regarding exclusive types, integers and booleans are encoded straightforwardly.
By definition of shared, values of type \gl{*}$t\exclusive$ and $t\shared$ have the same memory encoding.
% This is mirrored in the case for \gl{*}$t\exclusive$.
Structs, arrays, slices, and interfaces are encoded into custom mathematical types, carefully crafted with corresponding axioms.
Types within brackets \gl{[]} are type parameters of a mathematical type.
Regarding shared types, structs and arrays are encoded into special variants of the mathematical types used for exclusive types.
Since shared values are encoded as the memory underlying a data type, 
the mathematical types require a different axiomatization.
Lastly, all other shared types, referred to as $\otherwisetypevar$, are encoded to \gl{Ref}.

The encoding of operations on values mostly follows from our type encoding.
For instance, accessing a field \gl{e.f} is encoded as the tuple projection \gl{f($\llbracket$e$\rrbracket$)}. 
Shared values are often converted to exclusive values, for instance, whenever they are read from and not written to.
For values that are neither structs nor arrays, this conversion is straightforward,
since their exclusive value is stored in a single field.
Conversely, for structs and arrays, this conversion requires combining the values of multiple fields.

The encoding of access permissions can be derived from the type encoding and is as follows:
\begin{gobra}[numbers=none]
$\llbracket$ acc(&e: *$\otherwisetypevar$$\exclusive$) $\rrbracket$ $\encAs$ acc($\llbracket$e$\rrbracket$)
$\llbracket$ acc(&e: *struct{$\typevar_1, \dots, \typevar_n$}$\exclusive$) $\rrbracket$ $\encAs$ $\llbracket$acc(&e.$f_1$)$\rrbracket$ && $\dots$ && $\llbracket$acc(&e.$f_n$)$\rrbracket$
$\llbracket$ acc(&e: *[n]$\typevar$$\exclusive$]) $\rrbracket$ $\encAs$ $\forall$ i: Int $::$ 0 $\leq$ i $<$ n $\implies$ $\llbracket$acc(&e[i])$\rrbracket$
\end{gobra}
Permission to structs and arrays are encoded as permissions to all fields and indices, respectively. 
Otherwise, a permission is encoded as a field permission.

\section{Complete Search-and-Replace Example}\label{sec:full_examples}

\newcommand{\NL}{\linebreak \mbox{\textcolor{red}{$\hookrightarrow$}\space}}
Below, we present the full version of \figref{channel-perm-comm}.
Unlike the simplified excerpt, the full version specifies
and proves functional correctness of the \gl{searchAndReplace} method (\lineref{b_functional_correctness}).

As discussed in \secref{nutshell:concurrency}, we reason about the functional correctness of the method \gl{searchAndReplace} via the wait group: 
\begin{enumerate*}
  \item The main thread sends slice chunks through a channel to the workers.
  Together with each chunk, it also transfers the debt to process the chunk.
  The counter of the wait group keeps track of how many debts have to be payed (\gl{wg.Add} at \lineref{full_add}).
  \item When a worker has processed a chunk, it pays the debt back to the main thread and decreases the counter of the wait group (\gl{wg.Done} at \lineref{full_done}) accordingly.
  \item The main thread waits until all debts are payed off (\gl{wg.Wait} at \lineref{full_wait}) and then combines the results to prove that the entire slice is processed (\linerange{full_combine_start}{full_combine_end}).
\end{enumerate*}

The predicate \gl{messagePerm} (\lineref{full_message_perm}) contains all resources that are sent over the channel \gl{c}. 
In the simplified version, we only showed that the predicate contains the permissions to the slice chunk. 
In the full version, we show that the predicate also contains an instance of the predicate \gl{wg.UnitDebt}, capturing the debt of a worker.
The debt itself is specified through the predicate \gl{replacedPerm} (\lineref{full_replaced_perm}). 
To instantiate \gl{replacedPerm}, a worker has to provide the permissions to the chunk. Furthermore, all occurrences of \gl{x} have to be replaced with \gl{y} in the chunk.
The parameter \gl{chunk0} represents the original state of the chunk.
The partial application of \gl{replacedPerm} at \lineref{full_message_perm_unit_debt} fixes the argument with the initial state of the chunk.
The workers receive an instance of \gl{messagePerm} via a message sent on \lineref{full_send} and received on \lineref{full_receive}.
After the chunk is processed (\linerange{full_process_start}{full_process_end}),
\gl{replacedPerm} is instantiated (\lineref{full_fold_replaced_perm}).
The worker pays back the debt and then signals this with the call to \gl{wg.PayDebt} (\lineref{full_pay_debt}) and \gl{wg.Done} (\lineref{full_done}), respectively,
which transfers the instance of \gl{replacedPerm} to the main thread.
Because the debt has to be paid before the wait group is signaled,
it is guaranteed that all debts are paid back after \gl{wg.Wait} unblocks.
The proof annotations at \linerange{full_combine_start}{full_combine_end} combine the fact that all chunks are processed into the fact that the entire slice is processed and recover the permissions to the entire slice.

Besides the operations we described,
the code uses additional operations such as \gl{wg.Start} and \gl{wg.GenerateTokenAndDebt}.
They are required to establish the preconditions of the methods \gl{wg.Add}, \gl{wg.Done}, and \gl{wg.Wait}, but their purpose in this example is limited
to simple transformation of permissions. As such, we omit their discussion.

\begin{gobra}[breaklines=true, breakindent=5pt]
package pkg
import "sync"

pred messagePerm(wg *sync.WaitGroup, chunk []int, x, y int) { $\label{line:full_message_perm}$
  ($\forall$ i int $::$ 0 $\leq$ i $<$ len(chunk) $\implies$ acc(&chunk[i])) &&
  wg.UnitDebt(replacedPerm{toSeq(chunk), chunk, x, y}) $\label{line:full_message_perm_unit_debt}$
}

pred replacedPerm(chunk0 seq[int], chunk []int, x, y int) { $\label{line:full_replaced_perm}$
  len(chunk0) == len(chunk) &&
  ($\forall$ i int $::$ 0 $\leq$ i $<$ len(chunk) $\implies$ acc(&chunk[i])) &&
  $\forall$ i int $::$ 0 $\leq$ i $<$ len(chunk) $\implies$ §\NL§ chunk[i] == (chunk0[i] == x ? y : chunk0[i])
}

ghost
requires $\forall$ j int $::$ 0 $\leq$ j $<$ len(s) $\implies$ acc(&s[j], _)
ensures len(res) == len(s)
ensures $\forall$ j int $::$ 0 $\leq$ j $<$ len(s)$\implies$ s[j] == res[j]
pure func toSeq(s []int) (res seq[int]) { $\label{line:full_to_seq}$
  return len(s) == 0 ? seq[int]{} :
    toSeq(s[:len(s)-1]) ++ seq[int]{ s[len(s)-1] }
}

requires $\forall$ i int $::$ 0 $\leq$ i $<$ len(s) $\implies$ acc(&s[i])
ensures $\forall$ i int $::$ 0 $\leq$ i $<$ len(s) $\implies$ acc(&s[i])
ensures $\forall$ i int $::$ 0 $\leq$ i $<$ len(s) $\implies$ §\NL§ s[i] == (old(s[i]) == x ? y : old(s[i])) $\label{line:b_functional_correctness}$
func searchAndReplace(s []int, x, y int) {
  if len(s) == 0 {
      return
  }

  workers := 8
  workRange := 1000 

  // Initial state of the slice
  ghost s0 := toSeq(s)

  c := make(chan []int, 4)
  var wg@ sync.WaitGroup

  // predicate-name{$\dots$, _, $\dots$} is syntax for partial application
  ghost pr := messagePerm{&wg, _, x, y}
  // The second argument specifies the permission transferred 
  // from receiver to sender. Since the channel is asynchronous,
  // the argument is PredTrue, a predicate whose body is true.
  ghost c.Init(pr, PredTrue{}) $\label{line:b_chan_init}$
  ghost wg.Init()

  ghost seqs := seq[seq[int]]{}
  ghost pseqs := seq[pred()]{}

  // Spawn workers
  invariant acc(c.RecvChannel(), _)
  invariant c.RecvGivenPerm() == PredTrue{}
  invariant c.RecvGotPerm() == pr
  for i := 0; i != workers; i++ {
    go worker(c, &wg, x, y) $\label{line:b_spawn_threads}$
  }

  // Split slice into chunks, which are sent to workers
  invariant acc(c.SendChannel()) && c.SendGivenPerm() == pr
  invariant acc(wg.WaitGroupP(), 1/2) && !wg.WaitMode()
  invariant offset == 0 ? acc(wg.WaitGroupP(), 1/2) : §\NL§ acc(wg.WaitGroupStarted(), 1/2)
  invariant 0 $\leq$ offset $\leq$ len(s)
  invariant $\forall$ i int $::$ offset $\leq$ i $<$ len(s) $\implies$§\NL§ acc(&s[i]) && s[i] == s0[i]
  invariant offset != len(s) $\implies$§\NL§ offset == len(seqs) * workRange
  invariant offset == len(s) $\implies$ len(seqs) $>$ 0 &&
    len(s) == (len(seqs) - 1) * workRange +§\NL§ len(seqs[len(seqs) - 1])
  invariant $\forall$ i int $::$ 
    0 $\leq$ i $<$ len(seqs) - (offset == len(s) ? 1 : 0) $\implies$§\NL§len(seqs[i]) == workRange
  invariant $\forall$ i, j int $::$ 0 $\leq$ i $<$ len(seqs) &&
    0 $\leq$ j $<$ len(seqs[i]) $\implies$§\NL§ seqs[i][j] == s0[i * workRange + j]
  invariant len(pseqs) == len(seqs)
  invariant $\forall$ i int $::$ 0 $\leq$ i $<$ len(pseqs) $\implies$
    pseqs[i] == replacedPerm{seqs[i],§\NL§ s[i * workRange : i * workRange + len(seqs[i])], x, y}
  invariant $\forall$ i int $::$ 0 $\leq$ i $<$ len(pseqs) $\implies$§\NL§ wg.TokenById(pseqs[i], i)
  for offset := 0; offset != len(s); {
    nextOffset := offset + workRange
    if nextOffset > len(s) {
      nextOffset = len(s)
    }

    section := s[offset:nextOffset]
    assert $\forall$ i int $::$ 0 $\leq$ i $<$ len(s) $\implies$§\NL§ &section[i] == &s[i + offset]
    ghost s1 := toSeq(section)
    ghost wpr := replacedPerm{s1, section, x, y}
    wg.Add(1, 1, 2, PredTrue{}) $\label{line:full_add}$
    ghost if offset == 0 {
      wg.Start(1, 2, PredTrue{})
    }

    ghost wg.GenerateTokenAndDebt(wpr)
    fold wg.TokenById(wpr, len(pseqs))
    ghost seqs = seqs ++ seq[seq[int]]{ s1 }
    ghost pseqs = pseqs ++ seq[pred()]{ wpr }
    fold messagePerm{&wg, _, x, y}(section)
    c <-section $\label{line:full_send}$
    offset = nextOffset
  }

  ghost wg.SetWaitMode(1, 2, 1, 2)
  wg.Wait(1, 2, pseqs) $\label{line:full_wait}$
  ghost { $\label{line:full_combine_start}$
    invariant 0 $\leq$ i $\leq$ len(seqs)
    invariant $\forall$ j int $::$ i $\leq$ j $<$ len(seqs) $\implies$§\NL§ sync.InjEval(pseqs[j], j)
    invariant $\forall$ j int $::$
      0 $\leq$ j $<$ (i == len(seqs) ? len(s) : i * workRange) $\implies$§\NL§ acc(&s[j]) && s[j] == (s0[j] == x ? y : s0[j])
    for i := 0; i != len(pseqs); i++ { 
      unfold sync.InjEval(pseqs[i], i)
      low := i * workRange
      up := low + len(seqs[i])
      s1 := s[low:up]
      unfold replacedPerm{seqs[i], s1, x, y}()
      assert $\forall$ j int $::$ { &s[j] } low $\leq$ j $<$ up $\implies$§\NL§ &s[j] == &s1[j-low]
    }
  } $\label{line:full_combine_end}$
}

requires acc(c.RecvChannel(), _)
requires c.RecvGivenPerm() == PredTrue{}
requires c.RecvGotPerm() == messagePerm{wg, _, x, y}
func worker(c <-chan[]int, wg *sync.WaitGroup, x, y int) {
  fold acc(PredTrue{}(), 2/1)
  invariant PredTrue{}() && acc(c.RecvChannel(), _)
  invariant c.RecvGivenPerm() == PredTrue{}
  invariant c.RecvGotPerm() == messagePerm{wg, _, x, y}
  invariant ok $\implies$ messagePerm{wg, _, x, y}(s)
  for chunk, ok := <-c; ok; chunk, ok = <-c { $\label{line:full_receive}$
    unfold messagePerm{wg, _, x, y}(chunk) $\label{line:full_unfold_message_perm}$
    ghost chunk0 := toSeq(chunk) $\label{line:full_process_start}$
    invariant 0 $\leq$ i $\leq$ len(s)
    invariant $\forall$ j int $::$ 0 $\leq$ j $<$ len(s) $\implies$ acc(&chunk[j])
    invariant $\forall$ j int $::$ 0 $\leq$ j $<$ len(s) $\implies$§\NL§ chunk[j] == (chunk0[j] == x && j $<$ i ? y : chunk0[j])
    for i := 0; i != len(chunk); i++ {
      if chunk[i] == x {
        chunk[i] = y
      }
    } $\label{line:full_process_end}$
    fold replacedPerm{chunk0, chunk, x, y}() $\label{line:full_fold_replaced_perm}$
    ghost wg.PayDebt(replacedPerm{chunk0, chunk, x, y}) $\label{line:full_pay_debt}$
    wg.Done() // Same as wg.Add(-1) $\label{line:full_done}$
    fold PredTrue{}()
  }
}
\end{gobra}

\end{document}